# THE ROLE OF WEB OF SCIENCE PUBLICATIONS IN CHINA'S TENURE SYSTEM

Fei Shu[1,2], Wei Quan[3], Bikun Chen[4], Junping Qiu[1], Cassidy R. Sugimoto[5], Vincent Larivière[2,6]

[1] Chinese Academy of Science and Education Evaluation (CASSE), Hangzhou Dianzi University, Hangzhou, Zhejiang, China P.R.

[2] École de bibliothéconomie et des sciences de l'information, Université de Montréal, Montréal, Québec, Canada

[3] Educational Science Research Institute, Hunan University, Changsha, Hunan, China P.R.

[4] School of Economics and Management, Nanjing University of Science and Technology, Nanjing, Jiangsu, China P.R.

[5] School of Informatics and Computing, Indiana University Bloomington, Bloomington, Indiana, United States

[6] Observatoire des sciences et des technologies (OST), Centre interuniversitaire de recherche sur la science et la technologie (CIRST), Université du Québec à Montréal, Montréal, Québec, Canada

## ABSTRACT

Tenure provides a permanent position to faculty in higher education institutions. In North America, it is granted to those who have established a record of excellence in research, teaching and services in a limited period. However, in China, research excellence (represented by the number of Web of Science publications) is highly weighted in the tenure assessment compared to excellence in teaching and services, but this has never been systematically investigated. By analyzing the tenure assessment documents from Chinese universities, this study reveals the role of Web of Science publications in China's tenure system and presents the landscape of the tenure assessment process in Chinese higher education institutions.

## KEYWORDS

Web of Science, Tenure Assessment, China, University, Publication

## CORRESPONDING AUTHOR

Fei Shu
Tel: +1(514)9952608
Email: fei.shu@mail.mcgill.ca

## ACKNOWLEDGEMENTS

This study was partly supported by the Social Science and Humanities Research Council of Canada [75620190196].

# INTRODUCTION

The concept of tenure is central to North American higher education institutions. It provides a permanent position in the institution to faculty members who, after a certain number of years, have established a record of excellence in teaching, research and service. As per the 1940 report from the American Association of University Professors:

> "Tenure is a means to certain ends; specifically: (1) freedom of teaching and research and of extramural activities, and (2) a sufficient degree of economic security to make the profession attractive to men and women of ability. Freedom and economic security, hence, tenure, are indispensable to the success of an institution in fulfilling its obligations to its students and to society." (American Association of University Professors, 1940)

Since the 2000s, Chinese universities have adopted a tenure system, with the specific purpose of promoting research excellence (Zhang, 2014). While these policies have been known to put an important emphasis on journal publication—especially those indexed in international databases—no study has yet analyzed the content of tenure policies across a spectrum of Chinese institutions. This paper aims at contributing to a better understanding of publication requirements for tenure and promotion in China[1], with a focus on the role of one bibliometric database (Clarivate Analytics' Web of Science--WoS) in shaping these tenure policies.

## Tenure Assessment

Although research achievement, ability to attract funding, academic visibility, teaching excellence, and administrative or community service are all considered in the tenure process, the relative importance of each of these pillars of academe varies by institution (Chait, 2002). Prior to the 2000s, teaching was typically valued more than research and service, given that it accounted for more than half of the workload of tenure-track candidates (Diamond & Adam, 1998; Trower, 2002). Since the 2000s, however, research has replaced teaching and has been the most highly valued aspect in the tenure assessment process (Alperin et al., 2019; Gardner & Veliz, 2014; Harley, Acord, Earl-Novell, Lawrence, & King, 2010; McKiernan et al., 2019).

In this context, papers published in prestigious journals are regarded as evidence of research excellence (Gravestock & Greenleaf, 2008; Harley et al., 2010), which increases the pressure to publish in those venues (Alperin et al., 2019; Gardner & Veliz, 2014; Mcculloch, 2017; McKiernan et al., 2019; Rice & Sorcinelli, 2002). Several bibliometric indicators (e.g., JIF, h-index, etc.) are used to measure research excellence (Moher et al., 2018) although they are well known to be problematic, especially when applied at the individual researcher level (American Society for Cell Biology, 2013; Hicks, Wouters, Waltman, Rijcke, & Rafols, 2015). This is particularly true of journal-level indicators, which have been shown to poorly reflect the quality of individual papers (Archambault & Larivière, 2009; Larivière & Sugimoto, 2019; Lozano, Larivière, & Gingras, 2012; Seglen, 1997).

---

[1] In this study, China refers to mainland China, which is the geopolitical area under the direct jurisdiction of the People's Republic of China, excluding Hong Kong and Macau.

## Universities in China

As is the case in most countries (Larivière, Macaluso, Mongeon, Siler, & Sugimoto, 2018), universities account for the largest share of China's research output. They represent 82.6% of monographs and 74.4% of journal articles published in China (National Bureau of Statistics of China, 2016). There are 2,631 higher education institutions in China, including 1,236 universities offering undergraduate programs (Ministry of Education of China, 2017a). These universities can be stratified into three tiers (Figure 1) following the classification established by two national elite universities programs: Project 211 universities, Project 985 universities, and other universities.

Project 211 was initiated in 1995 by the Ministry of Education, aiming to construct 100 world-class universities in the twenty-first century (Ministry of Education of China, 2000). The Chinese government invested around 2.7 billion USD in this project, which offers preferential policies (access to best students and additional funding) to 112 admitted universities. These universities obtain 70% of the national research funding and matriculate 80% of doctoral students (Tang & Yang, 2008). Project 985 was announced by Zemin Jiang, former Chairman of the People's Republic of China, on May 4, 1998 to promote the development of a Chinese equivalent of the US Ivy League. It includes 39 universities; those are also in Project 211, which means that they can receive government funding from both national projects. Project 985 and Project 211 have been closed to new members since 2011, fixing the stratification of the Chinese university system into three tiers (Figure 1): Tier 1 (all universities admitted to Project 985), Tier 2 (all universities not admitted to Project 985 but admitted to Project 211) and Tier 3 (universities admitted neither to Project 985 nor to Project 211). Although a new national research program, "Double World-Class"[2], was initiated by the Chinese government in 2018, it was not in effect during the period of our investigation.

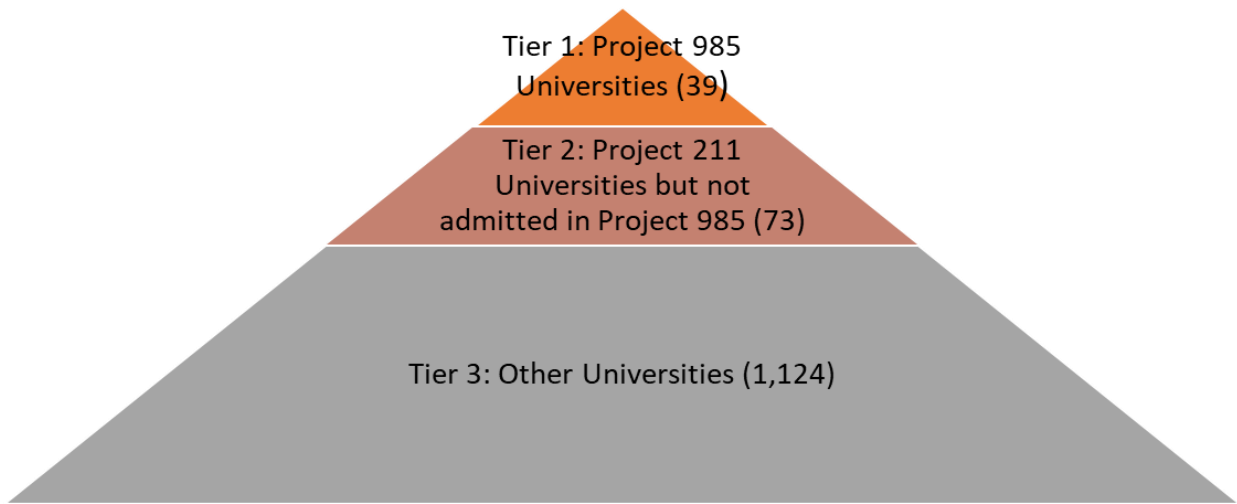

Figure 1. The Stratified System of Chinese Universities

---

[2]Double World-Class is a Chinese national research program to comprehensively develop a group of elite Chinese universities and individual university departments into world-class universities and disciplines by the end of 2050. (See http://www.gov.cn/zhengce/content/2015-11/05/content_10269.htm)

## China's Tenure System

Prior to the 1990s, all university professors in China—irrespective of rank—were considered as permanent staff of government affiliated institutions. Promotion through ranks was assessed by the Ministry of Education (MoE) or other ministries. However, such promotions were not associated with job security. In the 1990s, Chinese higher education institutions reformed this system and started to use contracts rather than permanent appointments. This was not, however, considered as a genuine tenure system as most contracts could be renewed with little or no evaluation (Lou, 2015).

The first tenure policy in China was introduced by Tsinghua University in 1994. It stipulated that an assistant professor would be dismissed if not promoted to associate professor within five years. Similar tenure policies have been adopted by other Chinese universities since the 2000s, generally with a six year tenure clock for new junior faculty (Zhang, 2014). As it is the case in most Western universities, Chinese scholars get tenure as they are promoted to Associate Professor. Despite being adopted by most Tier 1 and Tier 2 universities, as well as some Tier 3 universities, criteria for tenure vary. Previous studies have shown that Chinese institutions place excessive emphasis on the number of publications (Lou, 2015; Mohrman, Geng, & Wang, 2011; Quan, Chen, & Shu, 2017; L. Wang, 2013; Yang, 2014). However, little is known on 1) numbers of publications required for tenure; 2) how these publication requirements vary as a function of the "quality" of journals; and 3) if these policies differ from one university to another. Since tenure policies are recorded in internal (and often confidential) documents, criteria for assessment in China have never been systematically investigated. Using a sample of such assessment policies, this paper contributes to a better understanding of the tenure process in Chinese universities, and of how it focuses on scholarly publications. Although several other excellence criteria are considered in tenure policies (e.g., teaching and service), this paper focuses on their research component. Moreover, tenure requirements in terms of research vary, not only across different institutions, but also by discipline and stream (i.e., teaching stream, research stream, and general stream) within a single institution.

## METHODS

In order to present the landscape of the criteria for tenure assessment in China, we sampled 75 Chinese universities and investigated their tenure policies since 2010. Both stratified sampling and convenience sampling were used.

### Data Collection

First, considering the 3-tier pyramid hierarchy and regional differences among Chinese universities, we classified 1,243 Chinese universities into 21 categories by tier and region. Second, in order to create a representative sample, we retrieved tenure policies from universities in each category. Since tenure policies are often internal documents that may not be accessible, we had to select universities from each category based on data availability. We used the Chinese search engine Baidu to locate policies from the official websites of each selected university. Finally, a manual validation was conducted to ensure that the retrieved documents were official.

A total of 75 Chinese universities including 19 Tier 1 universities, 25 Tier 2 universities, and 31 Tier 3 universities from all seven regions were investigated (Table 1). While some universities may issue a general policy, other universities have different policies for each discipline and stream. On the whole, 256 tenure assessment policies were retrieved from these 75 universities: 23 universities contributed only one document, 14 universities contributed two, 15 universities contributed three while Wuhan university and Tongji University issued 18 and 17 documents respectively. As Figure 2 indicates, more than half of documents (150/256) contained general tenure requirements for all disciplines, while the remaining were only applied to specific disciplines. There were 162, 52 and 40 documents for the general stream, research stream and teaching stream, respectively.

Table 1. Number of sampled universities, by tier and region

| Region | Tier 1 | Tier 2 | Tier 3 | Total |
|---|---|---|---|---|
| North | 3 (10) | 4 (19) | 4 (178) | 11 (207) |
| Northeast | 2 (4) | 2 (7) | 1 (130) | 5 (141) |
| Northwest | 2 (4) | 3 (9) | 3 (94) | 8 (107) |
| Center | 3 (5) | 5 (7) | 5 (162) | 13 (174) |
| East | 5 (11) | 6 (20) | 8 (334) | 19 (365) |
| Southwest | 2 (3) | 2 (7) | 6 (127) | 10 (137) |
| South | 2 (2) | 3 (4) | 4 (99) | 9 (105) |
| Total | 19 (39) | 25 (73) | 31 (1,124) | 75 (1,236) |

Note: Numbers in brackets represent the total number of Chinese universities in each category.

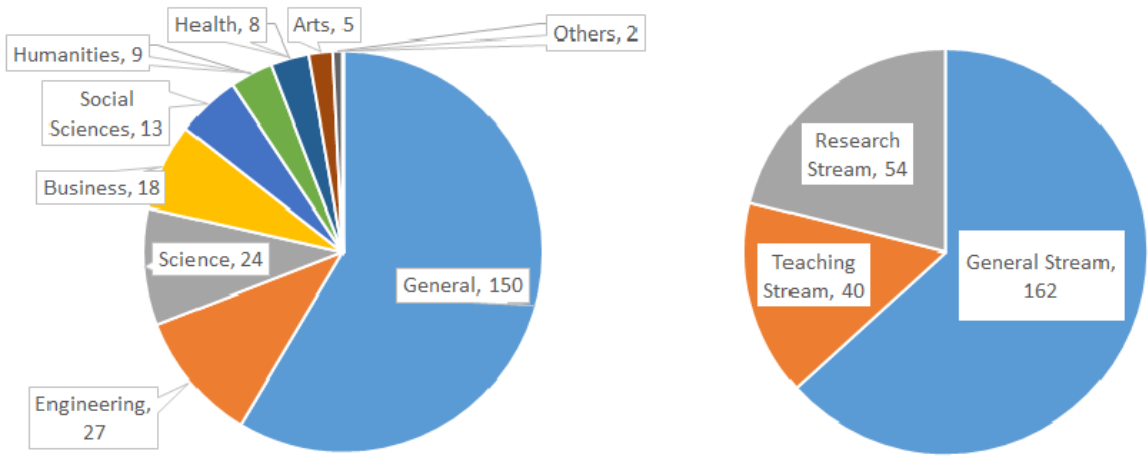

Figure 2. Distribution of tenure assessment documents by discipline (left panel) and stream (right panel)

Due to limited data availability, we did not use random sampling for our data collection, which is a limitation for this study. When comparing the research and development personnel (R&D personnel), number of scholarly articles, and research funding received between the sample and the population (see Table 2), we found that the means of these indicators from the sample Tier 1 universities were very close to those means from all Tier 1 universities, while the means from the sample Tier 2 universities were only a little higher than the means from all Tier 2 universities. Considering many Tier 3 universities are teaching colleges in which tenure policies are not adopted, the Tier 3 sample included many top Tier 3 universities so that the sample means were much higher than the average of all Tier 3 universities. We also did the one-sample T-test ($\alpha=0.05$) comparing the sample means with the population means to test whether the samples are representative. As Table 2 shows, we did not find any significant difference between sample and population in all three indicators in Tier 1 and Tier 2; significant difference was found between the Tier 3 sample and population. The T-test indicated that the Tier 1 and Tier 2 samples represented the population well while the Tier 3 sample only represented the top Tier 3 universities in this study.

Table 2 Comparison of Stats between the Sample Universities and All Universities in Average (2016)

|  | R&D personnel | Scholarly Paper | Research funding (USD: in millions) |
|---|---|---|---|
| Tier 1 Sample | 1,992 | 6,339 | 257.15 |
| All Tier 1 | 2,122 | 6,802 | 253.98 |
| P value (Tier 1) | 0.698 | 0.601 | 0.916 |
| Tier 2 Sample | 830 | 2,759 | 70.18 |
| All Tier 2 | 657 | 2,254 | 58.16 |
| P value (Tier 2) | 0.073 | 0.178 | 0.219 |
| Tier 3 Sample | 366 | 1,079 | 22.40 |
| All Tier 3 | 210 | 516 | 9.16 |
| P value (Tier 3) | 0.003 | 0.033 | 0.005 |

Source: Ministry of Education of China (2017b)

Data Analysis

Although each document contains various other criteria for tenure and promotion, only the means of tenure assessment in research was investigated, considering the scope of this study. As shown in Figure 3, all documents were analyzed to address the three following questions:

- Is publication a mandatory requirement for tenure and promotion?
- Is publishing in WoS-indexed journals a mandatory requirement for tenure and promotion?
- Are WoS papers benefiting from exceptional treatment (e.g., exemption from other requirements, or special promotion from assistant professor to full professor) in tenure assessment?

In addition to that information, we also retrieved data on requirements in terms of quantity of publications, impact factor of journals, authorship (i.e. author position) of publications, as well as requirements in terms of funding.

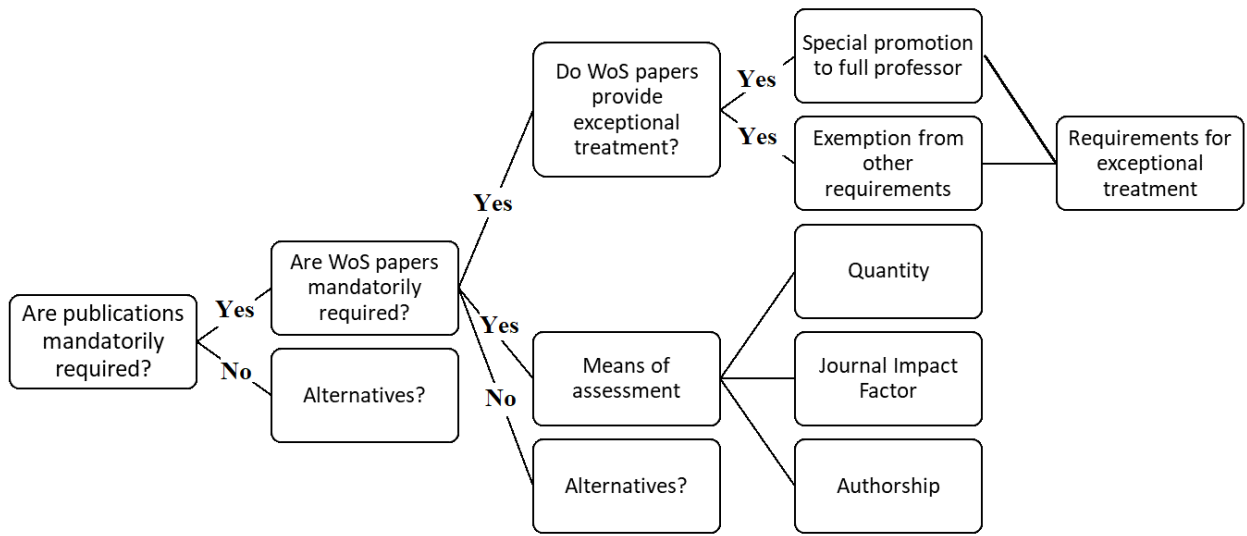

Figure 3. Procedure for data analysis

## RESULTS

After investigating 256 tenure assessment documents from 75 Chinese universities, we found that scholarly publications, and especially WoS-indexed papers, were a mandatory requirement for tenure and promotion in most universities. Moreover, in some cases, tenure-track candidates could get exceptional treatment (e.g., direct promotion from assistant professor to full professor with tenure) if they had reached a certain number of WoS-indexed papers.

### Publication as Mandatory Requirement

Among the 256 tenure assessment documents, 88.7% (227/256) indicate that having a publication record is a mandatory requirement for tenure and promotion. Quite surprisingly, an even higher proportion (95%) of tenure assessment for the teaching stream (38/40) also require publications. Moreover, for 23 out of the 29 policies that did not require scholarly publications, other requirements in research such as patents, scholarships, grants and awards were required, which, in many cases, are obtained based on a strong publication record. Such publication requirements were relatively homogeneous across disciplines or universities, as well as tiers and regions. It is worth mentioning that, in general, only specific authorship positions are recognized in tenure assessment policies: indeed, most universities (89.5%) only count publications in which candidates are first or corresponding author.

Only six documents from four distinct universities show that tenure-track candidates could are able to get tenure without considering research. In those cases, candidates need to demonstrate great achievement in teaching. East China University of Science and Technology (Tier 2) requires that tenure-track candidates in the teaching stream make a significant contribution to the curriculum. At Peking University (Tier 1), tenure-track candidates in the teaching stream should have 9 years of teaching experience and have completed a funded teaching project at provincial or ministry level. Without publications, tenure-track candidates in social science at Ocean University of China (Tier 1) must hold a national teaching award. For tenure-track candidates from all three streams at Nanchang University (Tier 2), teaching awards can be a replacement for publications for tenure assessment.

## WoS Paper as a Mandatory Requirement

Papers indexed in the WoS have been shown to have a special status in China in terms of monetary rewards (Quan et al., 2017). This is also observed for tenure assessment: WoS-indexed papers are often considered as a mandatory requirement for tenure in several Chinese universities. As shown in Table 3, 17.6% of all tenure assessment documents for which a publication record is mandatory (40/227) indicate that tenure-track candidates need to publish WoS papers for tenure and promotion. This percentage varies strongly by tier and discipline. Only 1 out of the 69 tenure assessment documents (1.4%) issued by Tier 3 universities requires WoS papers for tenure and promotion while 38.2% of documents issued by Tier 1 universities have the same requirement. There is no WoS paper requirement for tenure-track candidates in Arts and Humanities (0/13) but the majority of tenure assessment documents require WoS papers for tenure and promotion in Health (85.7%, 6/7) and Science (54.5%, 12/22), especially for Tier 1 universities.

Table 3. For the subset of papers where publication is a mandatory requirement: percentage of tenure assessment documents in which WoS paper is mandatory, by tier and discipline

| Discipline | Tier 1 | Tier 2 | Tier 3 | All Universities |
| --- | --- | --- | --- | --- |
| Arts | 0.0% | - | 0.0% | 0.0% |
| Humanities | 0.0% | 0.0% | - | 0.0% |
| Social Sciences | 16.7% | 11.1% | 0.0% | 8.2% |
| Business | 22.2% | 12.5% | - | 17.6% |
| Health | 85.7% | - | - | 85.7% |
| Engineering | 43.8% | 22.2% | 0.0% | 34.6% |
| Science | 83.3% | 22.2% | 0.0% | 54.5% |
| General Natural Science | 50.0% | 7.7% | 6.7% | 12.5% |
| General All Disciplines | 0.0% | 9.1% | 0.0% | 3.3% |
| Total | 38.2% | 12.2% | 1.4% | 17.6% |

Results also show that 26.1% of tenure assessment documents (12/46) require WoS papers for tenure-track candidates in the research stream while 13.9% of documents have the same requirement for candidates in the teaching stream. Tenure-track candidates in Natural Sciences in the teaching stream from Wuhan University (Tier 1) and Northwestern Polytechnical University need to publish WoS papers to get promoted.

Although WoS papers are not a mandatory requirement in 187 tenure assessment documents, they still help tenure-track candidates fulfill research requirements for tenure and promotion. According to these 187 documents, in addition to WoS papers, the tenure assessment recognizes monographs (7/187) and journal articles (180/187) indexed by other bibliometric databases, e.g. Chinese Science Citation Database (CSCD), Chinese Social Science Citation Index (CSSCI), Engineering Index (EI).

Among the 40 tenure policies in which WoS papers are a mandatory requirement, 22 only count the number of WoS papers, requiring from between one and three papers as a minimum, while 18 also consider the impact factor of the journals in which they are published. For example, tenure-track candidates at Southwest Jiaotong University (Tier 2) need to publish three WoS papers in journals that are included in Quartile 1 or 2 of the Journal Citation Report. Southeast University (Tier 1) requires that tenure-track candidates in Health and Science need to publish at least three WoS papers or that the sum of the JIFs of the journals in which the papers are published is over 10 (e.g., one PNAS paper). Candidates in Health and Science at Tongji University (Tier 1) have more options to fulfill publication requirements: five WoS Papers, or three WoS papers published in the top 2 journals (based on JIF) in their disciplines, or one paper in *Nature* or *Science*, or one paper that was cited more than 30 times.

### WoS Papers and Exceptional Treatment

Candidates with strong publication records can also benefit from exceptional treatments—especially when they fail to meet other requirements. More specifically, 73 out of 256 tenure assessment documents (28.5%) show that candidates could be exempt from other requirements if they have a very strong publication record; they can even be promoted to full professor in some special cases. For example, in addition to fulfilling other requirements (e.g., teaching, grants, awards), candidates in Health and Science from Southeast University (Tier 1) need to publish at least three WoS papers, or one or two WoS papers but the sum of the JIF of the journals in which they were published must be higher than 10. If these numbers (i.e., 3 and 10) increase to 8 and 20 respectively, candidates can get tenure without fulfilling any other requirements; if these numbers increase to 18 and 35 respectively, or they publish a paper in *Nature* or *Science*, they can directly become full professors.

As shown in Table 4, over one third of policies issued by Tier 1 (34.4%, 31/90) and Tier 3 universities (34.2%, 25/73) offer exceptional treatment for WoS papers while only 18.3% of documents issued by Tier 2 universities (17/93) contain similar policies. Exceptional treatments appear in 62.5% of tenure assessment documents in Health and more than 30% of documents in either Natural Science or Social Science. An interesting finding is that 30% of tenure assessment documents (12/40) for the teaching stream also contain such exceptional treatment. This means that tenure-track candidates in the teaching stream can get tenure without any teaching achievement.

While 15 out of 73 tenure assessment documents (20.5%) that include exceptional treatment show that tenure-track candidates with a very strong publication record can be exempt from some of the other requirements (e.g., length of the probatianary period, education background, etc.), the rest (79.5%, 58/73) indicate that a strong publication record exempts candidates from all other requirements. In other words, in these cases, prolific scholarly publication activity remains the only requirement for tenure.

Table 4. Percentage of tenure assessment documents containing exceptional treatment by tier and discipline

| Discipline | Tier 1 | Tier 2 | Tier 3 | All Universities |
|---|---|---|---|---|
| Arts | 0.0% | - | 0.0% | 0.0% |
| Humanities | 16.7% | 0.0% | - | 11.1% |
| Social Sciences | 50.0% | 27.8% | 30.0% | 34.6% |
| Business | 30.0% | 0.0% | - | 16.7% |
| Health | 62.5% | - | - | 62.5% |
| Engineering | 38.9% | 22.2% | 0.0% | 32.1% |
| Science | 50.0% | 22.2% | 0.0% | 37.5% |
| General Natural Science | 20.0% | 23.1% | 56.3% | 38.2% |
| General All Disciplines | 0.0% | 15.2% | 29.4% | 19.2% |
| Total | 34.4% | 18.3% | 34.2% | 28.5% |

Furthermore, with such prolific publication records, candidates can be promoted to full professors at the same time as getting tenure. For instance, 37 tenure assessment documents show that this special promotion to full professor is available. The requirements of special promotion vary, but they are generally similar to the tenure requirements of Southeast University described above. Regardless of whether WoS papers are mandatory for tenure, tenure-track candidates could obtain direct promotion to full professor with an excellent publication record. Figure 4 presents another example of how tenure-track candidates (research stream) in Social Sciences at Wuhan University (Tier 1) can get tenure. They can teach one undergraduate course and a minimum 24 hours teaching per year, fulfill three research requirements (that is, 6 CSSCI papers, 1 grant from either National Science Foundations of China or National Social Science Foundations of China, and 1 monograph or book) to get tenure. However, if the candidate publishes a WoS paper, she or he could be exempt from all other tenure requirements; if she or he could publish a paper in *Nature* or *Science*, the candidate could win the special promotion and be promoted to full professor directly.

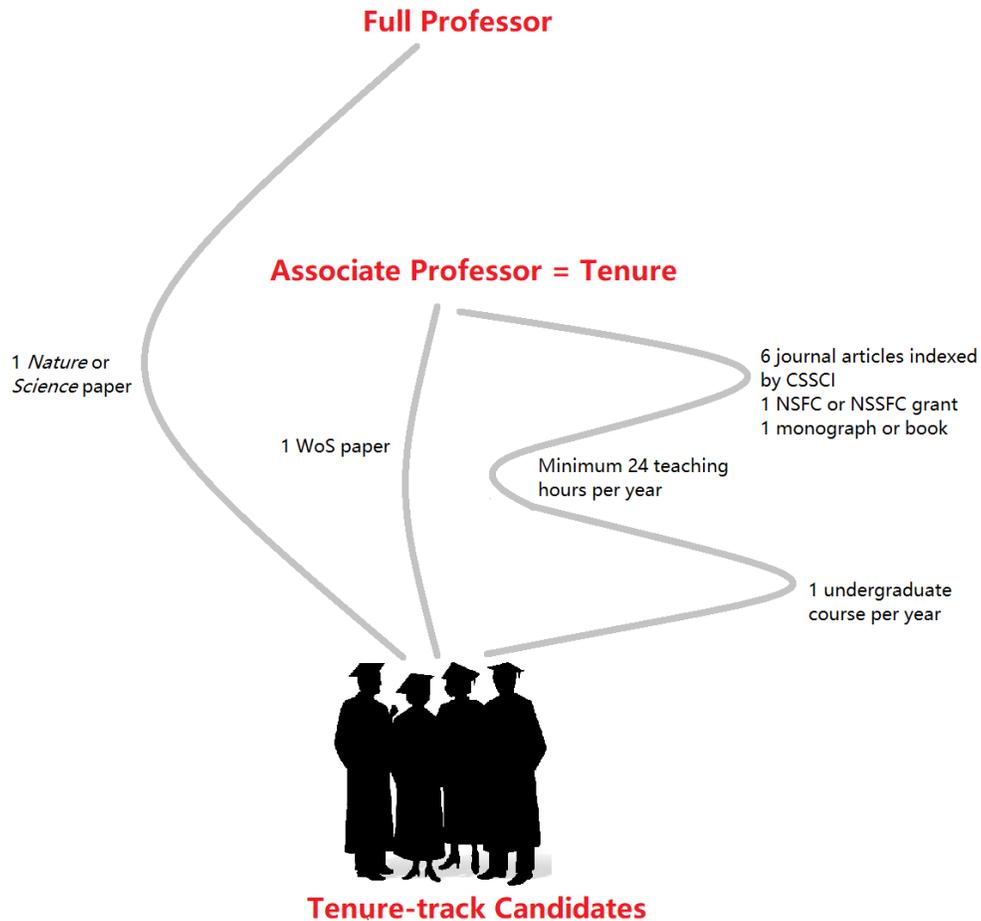

Figure 4. Example of the road to tenure at Wuhan University (Social Sciences, Research Stream)

## DISCUSSION AND CONCLUSION

Since the 1980s, WoS papers have been used in China to evaluate research performance and to increase the international visibility of Chinese research (Gong & Qu, 2010; Quan et al., 2017). It has previously been shown that Chinese scholars, especially in the Natural Sciences, are required to publish WoS papers to get tenure and promotion (Lou, 2015; Mohrman et al., 2011; Quan et al., 2017; L. Wang, 2013; Y. Wang & Li, 2015; Yang, 2014)—a finding that has been confirmed by this study. Publishing papers indexed in the WoS is the mandatory requirement in most tenure assessment documents for tenure-track candidates in the Natural Sciences, even for candidates who are in the teaching stream. Publishing WoS papers can even, in some cases, exempt candidates from other tenure requirements, or provide a direct promotion to full professor.

Although the purpose of the tenure system is to defend the principle of academic freedom—in which scholars are free to hold and examine a variety of views, both in teaching and research—our results suggest that, in the case of China, tenure is mostly seen as an incentive to increase the research productivity of assistant professors. As this study indicates, publication record is at the heart of tenure assessment policies, which disfavor candidates who are not strong in research. Such trends are not entirely specific to Chinese universities, however. As shown by Alperin et al. (2019), a high percentage of North American universities (40%) also discuss bibliometric indicators in their tenure policies, and research is generally more highly valued than teaching. However, what makes the Chinese case specific is that tenure could be obtained *exclusively* on bibliometric indicators, and that this path to tenure is formalized in university policies. Indeed, although teaching experience and other requirements are still included in tenure assessment documents, they are meaningless when exceptional treatments are possible; and these exceptional treatments are only based on scholarly publications. An excellent publication record (i.e., publishing a lot of WoS papers) help candidates get tenure or direct promotion to full professor, regardless of teaching and service. On the other hand, excellent lecturers who are appreciated by their students may not get tenure because of their lack of publications (Zhao & Pan, 2014).

Tenure criteria create incentives. They influence what candidates focus their attention on and the activities they choose to pursue (Harley et al., 2010). In China, to satisfy these criteria for tenure assessment, tenure-track candidates have to spend most of their work time building their publication portfolio—keeping in mind indicators such as the impact factor—rather than developing their teaching activities and service to the community, which are also core functions of university professors. Moreover, Chinese universities do not offer non-tenure-track positions (e.g., Lecturer, Adjunct Professor, etc.) for teaching. As a result, one may argue that over the past 20 years, tenure policies have helped develop research excellence of China's higher education, but at the expense of the quality of teaching and service.

Our analysis does have some limitations. First and foremost, due to the availability of tenure policies, we could not use random sampling and had to rely on convenience sampling for our data collection. This may influence results: available tenure policies may rely more (or less) on research than those that are confidential. Second, although this study presents a landscape of the tenure policies in China, we did not investigate whether there is a relationship, at the level of institutions, between such policies and research performance (or scientific misconduct); this could be explored by future research.